# Neural Avalanches at the Critical Point between Replay and Non-Replay of Spatiotemporal Patterns

Silvia Scarpetta[1,2]*, Antonio de Candia[3,4,5]

1 Dipartimento di Fisica "E. R. Caianiello", Università di Salerno, Fisciano (SA), Italy, 2 INFN Gr. Coll. di Salerno, Fisciano (SA), Italy, 3 Dipartimento di Fisica, Università di Napoli Federico II, Napoli, Italy, 4 CNR-SPIN, Sezione di Napoli, Napoli, Italy, 5 INFN, Sezione di Napoli, Complesso Universitario di Monte S. Angelo, Naples, Italy

## Abstract

We model spontaneous cortical activity with a network of coupled spiking units, in which multiple spatio-temporal patterns are stored as dynamical attractors. We introduce an order parameter, which measures the overlap (similarity) between the activity of the network and the stored patterns. We find that, depending on the excitability of the network, different working regimes are possible. For high excitability, the dynamical attractors are stable, and a collective activity that replays one of the stored patterns emerges spontaneously, while for low excitability, no replay is induced. Between these two regimes, there is a critical region in which the dynamical attractors are unstable, and intermittent short replays are induced by noise. At the critical spiking threshold, the order parameter goes from zero to one, and its fluctuations are maximized, as expected for a phase transition (and as observed in recent experimental results in the brain). Notably, in this critical region, the avalanche size and duration distributions follow power laws. Critical exponents are consistent with a scaling relationship observed recently in neural avalanches measurements. In conclusion, our simple model suggests that avalanche power laws in cortical spontaneous activity may be the effect of a network at the critical point between the replay and non-replay of spatio-temporal patterns.

**Funding:** This work was supported by university public funds from the University of Naples "Federico II" and University of Salerno. The funders had no role in study design, data collection and analysis, decision to publish, or preparation of the manuscript.

**Competing Interests:** The authors have declared that no competing interests exist.

* E-mail: silvia@sa.infn.it

## Introduction

Recently, many experimental results have supported the idea that the brain operates near a critical point [1–5], as reflected by the power laws of avalanche size distributions and maximization of fluctuations. Several models have been proposed as explanations for the power law distributions that emerge in spontaneous cortical activity [5,6]. Models based on branching processes [4] and on self-organized criticality [7–10] are the most relevant.

However, there are additional features of neuronal avalanches that are not captured in these models, such as the stable recurrence of particular spatio-temporal patterns and the conditions under which these precise and diverse patterns can be retrieved [4]. Indeed, neuronal avalanches are highly repeatable and can be clustered into statistically significant families of activity patterns that satisfy several requirements of a memory substrate [11–13].

In many areas of the brain having different brain functionality, repeatable precise spatio-temporal patterns of spikes seem to play a crucial role in the coding and storage of information. Many in vitro [14,15] and in vivo [16–18] studies have demonstrated that cortical spontaneous activity occurs in precise spatio-temporal patterns, which often reflect the activity produced by external or sensory inputs. The temporally structured replay of spatio-temporal patterns has been observed to occur, both in the cortex and hippocampus, during sleep [16,19,20] and in the awake state [21–24], and it has been hypothesized that this replay may subserve memory consolidation.

Further evidence on the central role played by precise phase-coded spatio-temporal patterns comes from the experiments on spike-phase coding of natural stimuli in the auditory and visual primary cortices [25,26] and from experiments on the short-term memory of multiple objects in the prefrontal cortices of monkeys [27].

Previous studies have separately addressed the topics of phase-coded memory storage and neuronal avalanches, but our work is the first to show how these ideas converge in a single cortical model. We study a network of leaky integrate-and-fire (LIF) neurons, whose synaptic connections are designed with a rule based on spike-timing-dependent plasticity (STDP). The network works as an associative memory of phase-coded spatio-temporal patterns, whose storage capacity has been studied in [28].

In this paper, we show that if the excitability of the model is tuned to be at the critical point of a phase transition, between the successful persistent replay of stored patterns and non-replay, then the spontaneous activity is characterized by power laws in avalanche size and duration distributions, critical exponents consistent with scaling relations, and maximization of order parameter fluctuations, as observed in many experiments.

In the cortex, the emergence of power law distributions of avalanche sizes depends on an optimal concentration of dopamine [13] and on the balance of excitation and inhibition [5,29], suggesting that particular parameters must be appropriately tuned. This may suggest that the cortex operates near the critical point of a phase transition, characterized by a critical value of excitability.

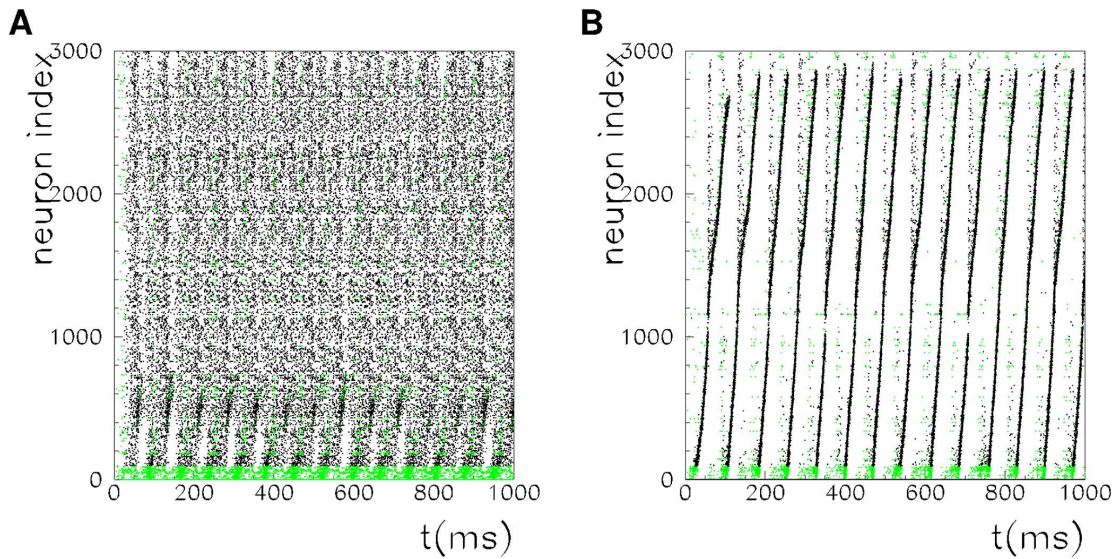

**Figure 1. Spontaneous dynamics of a network of $N=3000$ units, with thresholds $\Theta_1=0.8$ and $\Theta_2=2.7$.** The same activity is shown with neurons sorted according to the phase $\phi_i^\mu$ with $\mu=1$ in A and $\mu=2$ in B. In this supercritical regime, a persistent reactivation of one of the patterns (randomly chosen) emerges, in this case, pattern $\mu=2$, as shown by the regular behavior in B. The behavior in A shows that pattern $\mu=1$ is not continuously replayed.
doi:10.1371/journal.pone.0064162.g001

This idea is also supported by experimental results, showing a high value of fluctuations [5] in correspondence with power law distributions, as expected at a critical point of a phase transition.

Notably, also large-scale fMRI analysis [2] demonstrates that the resting brain spends most of the time near the critical point of a second-order transition and exhibits avalanches of activity ruled by the same dynamical and statistical properties described previously for neuronal events at smaller scales.

## Results

We model cortical activity with a coupled network of LIF units, using the Spike Response Model formulation [30,31]. The postsynaptic membrane potential of each neuron $i$ is given by a Possonian noise $\eta_i(t)$ plus the sum, weighted by synaptic connections $J_{ij}$, of the response kernels to incoming spikes of presynaptic units. In terms of in vitro cortical cultures, the source of noise that we model is related to the spontaneous neurotransmitter release at individual synapses, as well as other sources of inhomogeneity and randomness that determine an irregular background synaptic noise in vitro.

Connectivity governs the collective spontaneous dynamics. Connections $J_{ij}$ between units are designed via the learning rule, inspired by the STDP, previously introduced in [28,32–34]. The importance of spike timing for synaptic plasticity has been observed in many brain areas [35,36], and its computational relevance has been analysed from different point of views [28,35,37–39].

While in [28] we studied the dynamics induced by an external cue stimulation and showed that a cue with few spikes with the proper phase relationships is able to induce the replay of the stored pattern in a proper region of parameters, here we study the spontaneous dynamics in the absence of any cue external stimulation in a noisy environment. Moreover, while in [28] a unique value of the spiking threshold $\Theta^i$ is used for all units, here we model the heterogeneity of the neurons excitability, using two values of $\Theta^i$, a low threshold $\Theta_1$ for a small number $N_1<N$ of units, and a higher threshold $\Theta_2$ for the other $N_2=N-N_1$ units. Indeed, as shown in many raster plots of *in-vitro* spontaneous dynamics with neural avalanches, there is often a small subset of units which have a higher spiking rate than the others. These are modeled here by the lower-threshold units, that are more sensible to noise. If some of these units have consecutive phases in one of the stored patterns, then the replay of the pattern is more easily triggered by noise. The value of the threshold $\Theta_1$ determines mainly the probability of activation of the replay of patterns, while the threshold $\Theta_2$ of the majority of the units will determine the duration of the replay, and the distribution of avalanches in the critical regime. For this reason we here fix the concentration of lower-threshold units and the value of their threshold, and study the behavior of the network as a function of $\Theta_2$. We show indeed that, in the absence of any external stimulation, noise is able to induce an intermittent replay of the stored phase patterns at some critical value of spiking threshold $\Theta_2$, and a permanent replay of one of the patterns at lower values of spiking threshold $\Theta_2$.

Figures 1 and 2 show the spontaneous dynamics of a network of $N=3000$ units, $P=2$ stored patterns, spiking threshold $\Theta_1=0.8$, and spiking threshold $\Theta_2=2.7$ and 3.0, respectively, in Figs. 1 and 2. The spikes of low-threshold neurons (units with $\Theta_1$) are plotted in green, while those of high-threshold neurons (units with $\Theta_2$) are plotted in black. As evidence of the replay of different patterns, we show the raster plot of the network dynamics with different sortings on the vertical axes. In Figs. 1a and 2a, neurons are sorted according to increasing values of the phases in pattern $\mu=1$, while in Figs. 1b and 2b, they are sorted according to increasing values of the phases in pattern $\mu=2$.

At a low value of the threshold $\Theta_2$, namely, $\Theta_2=2.7$, the first pattern that is replayed (randomly chosen by the noise) goes on for a very long, apparently infinite, time, as shown in Fig. 1b, where only pattern $\mu=2$ is replayed. The same sequence of spikes, shown in Fig. 1a, does not reveal any long lasting ordered sequence, showing that pattern $\mu=1$ is not continuously replayed during the same interval of time. Note that the noise triggers some short replays of pattern $\mu=1$, that however do not survive due to the intereference with the permanent replay of pattern $\mu=2$ that is going on. This is confirmed by the order parameter, introduced in the next subsection, that is of order $1/\sqrt{N}$ in this case for pattern $\mu=1$, and of order one for pattern $\mu=2$. With this connectivity and this value of threshold the dynamics of the network tends to be oscillatory, with the same phase relationship of one of the stored pattern, but with an oscillation frequency $\nu$ different from the stored frequency $\nu^\mu$, namely the replay dynamics is faster then the one of the stored patterns (see also [28]).

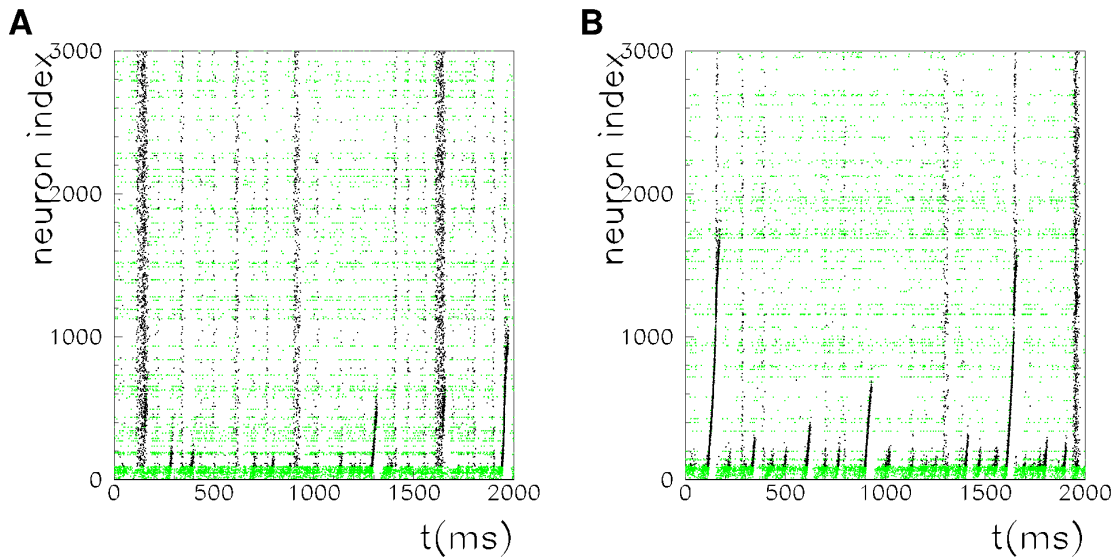

**Figure 2. Spontaneous dynamics of a network of $N=3000$ units, with thresholds $\Theta_1=0.8$ and $\Theta_2=3.0$.** Neurons are sorted as in Fig. 1. In this critical regime, an intermittent spontaneous replay of both the patterns is observed.
doi:10.1371/journal.pone.0064162.g002

At a little higher value of $\Theta_2$, namely, $\Theta_2=3.0$, the behavior is very different (see Figs. 2a and 2b). It can be seed that, from time to time, there is a short transient replay of one of the two patterns. When pattern $\mu=1$ is replayed, a short sorted sequence of spikes appears in Fig. 2a, while when pattern $\mu=2$ is retrieved, a short sorted sequence of spikes appears in Fig. 2b. Note that, when the pattern $\mu=1$ is replayed, a chaotic burst of spikes appears in Fig. 2b that is sorted according to the other pattern $\mu=2$ and vice versa.

At still higher values of the threshold (not shown), neither of the patterns is replayed for a time long enough to be distinguishable from noise.

### The Order Parameter and the Phase Transition

To measure the success of the replay, we introduce a quantity that estimates the overlap between the network collective activity during the spontaneous dynamics and the stored phase-coded pattern. This quantity is maximal (equal to one) when collective activity is periodic, as in Fig. 1b, and the ordering of spiking times coincides with that of one of the stored patterns, and is of order $\simeq 1/\sqrt{N}$ when the spike timings are uncorrelated with the stored ones. The overlap $Q^\mu(T^w)$ is defined as the average of the time-dependent quantity $q(t,t+T^W)$, namely

$$Q^\mu(T^w) = \langle |q(t,t+T^w)| \rangle$$

where

$$q(t,t+T^w) = \frac{1}{N_s} \sum_{\substack{j=1,\ldots,N \\ t<t_j^*<t+T^w}} e^{-i2\pi t_j^*/T^w} e^{i2\pi t_j^\mu/T^\mu} \quad (1)$$

$t_j^\mu$ are the spike times in the stored phase-coded pattern $\mu$ with period $T^\mu$, $t_j^*$ are the spike times of neuron $j$ during the spontaneous collective dynamics, $T^w$ is a "probe" window of time, the average $\langle \cdots \rangle$ is done on the starting time $t$ of the window, and $N_s$ is the number of spikes in the time interval $T^w$. The fluctuations of the overlap are given by

$$\sigma^2(Q^\mu) = N\left[\langle |q(t,T^w)|^2\rangle - \langle |q(t,T^w)|\rangle^2\right]. \quad (2)$$

As the overlap $q(t,t+T^w)$ is an intensive quantity, that is it does not depend on the number $N$ of neurons when $N$ is large, we expect that its fluctuations are of order $1/N$, and therefore add a factor $N$ in Eq. (2).

In Figs. 3a and 3b, we show the overlap and its fluctuations, respectively, as a function of $T^w$ for $N=3000$, $\Theta_1=0.8$, and three values of $\Theta_2$, namely, $\Theta_2=2.7, 3.0, 3.3$. We see that the overlap always has a maximum at some value of $T^w$. This corresponds (approximately) to the period of the pattern during replay, both when the pattern is replayed continuously, as in Fig. 1b, and when there are short and incoherent partial segments of different patterns, as in Figs. 2a and 2b. We therefore define the order parameter $m^\mu$ as the maximum of the overlap as a function of $T^w$,

$$m^\mu = \max_{T^w} Q^\mu(T^w). \quad (3)$$

This definition works also when the periodic pattern is not replayed continuously, and short replays are hidden in a nonperiodic spike train, such as here and in many experimental situations. The fluctuations $\sigma^2(m^\mu)$ of the order parameter are defined as the fluctuations of the $T^w$-dependent overlap, at the same $T^w$ where the overlap has its maximum.

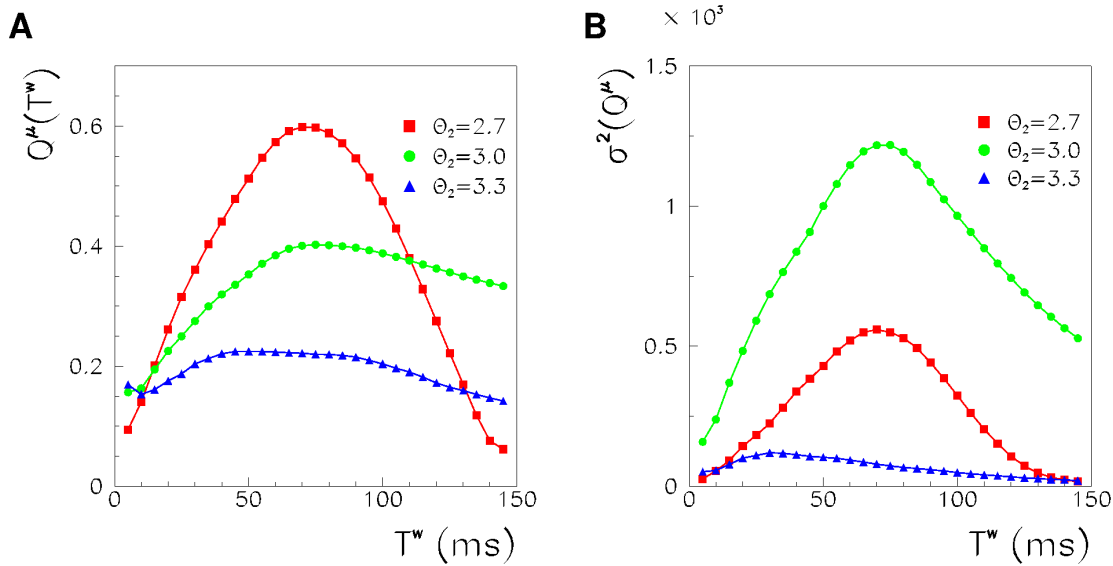

**Figure 3. Overlap (A) and its fluctuations (B) as a function of the chosen window $T^w$, with $N = 3000$, $\Theta_1 = 0.8$, and $\Theta_2 = 2.7$ (squares), 3.0 (circles) and 3.3 (triangles).** Note that, while overlap increases when the spiking threshold $\Theta_2$ decreases, the fluctuations are larger at the critical value of the threshold.
doi:10.1371/journal.pone.0064162.g003

In Figs. 4a and 4b, the behavior of the order parameter and its fluctuations, as a function of the spiking threshold $\Theta_2$ and for different sizes of the network, are shown. At a low-spiking threshold, the order parameter is high, and fluctuations are low, indicating that, as shown in Fig. 1a, the noise is able to initiate a successful long-lasting replay of the stored pattern. At high thresholds, both the order parameter and its fluctuations are low. At the critical point between the two regimes, the fluctuations of the order parameter are maximized, and the maximum seems to diverge at the transition with the size of the network, as happens in a continuous phase transition. This suggests that there is not a defined timescale of the replayed segments but rather a scale-free power law distribution. We therefore, in the next Section, study the distributions of the durations and sizes of the replayed segments. Notably the phase transition that we find here is not a thermodynamical phase transition but a non-equilibrium phase transition, defined using a dynamical order parameter that is an extension of the Hopfield order parameter but for phase-coded dynamical states.

### The Critical Point and Neural Avalanches

In order to characterize the noise-induced collective dynamics near the critical point, we study the interspike-interval statistics and the sizes and durations of the avalanches of spikes.

The distribution of interspike intervals (ISI) among consecutive spikes over all of the network is shown in Fig. 5 for $N = 20000$ and spiking threshold $\Theta_2 = 2.5, 3.0, 3.7$. We note that, while at high- and low-spiking thresholds the network ISI distribution is well described by an exponential, only at the critical threshold is the network ISI clearly not exponential. The distributions at $\Theta_2 = 2.5$ and 3.7 are well described by the exponential fit $e^{-\tau/\tau_0}$, with $\tau_0 = 0.003$ ms and $\tau_0 = 0.015$ ms, respectively. On the other hand, the critical distribution at $\Theta_2 = 3.0$ starts with an exponential with $\tau_0 = 0.006$ ms but at $\tau = 0.03$ ms deviates strongly from the exponential behavior. This makes a strong link between the criticality observed in terms of the order parameter and the spiking dynamics characterized in terms of the network ISI. The coexistence of many time scales at the critical point is revealed also in the shape of the network ISI distribution.

To study the network dynamics in terms of avalanches of activity, we define an avalanche as a sequence of spikes preceded and succeeded by a time interval of length at least $\tau_{min}$ without any spikes. The value of $\tau_{min}$ has been chosen looking at the network ISI distributions as a value greater than the short timescale of ISIs within an event but less than the timescale of the longer quiescent periods, which are not distributed exponentially. Therefore, we take a value of $\tau_{min} = 0.03$ ms as the time at which the ISI distribution at the critical point deviates from the initial exponential behavior.

For each avalanche, we measure its duration $T$ in ms and its size $s$ defined as the total number of spikes within the avalanche. Figure 6a shows the size distributions at the three different spiking thresholds. At the critical point ($\Theta_2 = 3.0$), the size distribution is a power law, and the fit $P(s) \propto s^{-\beta}$ gives an exponent $\beta = 1.55$.

Figure 6b shows the duration distribution for the three regimes, showing that at the critical point $\Theta_2 = 3.0$ the duration distribution approaches a power law, well fitted by $P(T) \propto T^{-\alpha}$ with $\alpha = 1.63$. Note that, for the values $\Theta_2 = 3$ and $3.7$, when there is not a permanent replay of one pattern, there is an initial exponential regime of the size and duration distributions. This is due to the fact that only a small fraction of the avalanches of low threshold units are able to trigger a larger avalanche of high threshold units. The remaining majority of avalanches of low threshold units have an exponential distribution with a small characteristic size and duration, independent from the value of $\Theta_2$.

Finally, Fig. 7 shows the average size $\langle s \rangle(T)$ of the avalanche of duration $T$, as a function of duration $T$. Again, the function approaches a power law, with exponent $k = 1.14$. Note that the critical exponents satisfy the scaling relation $\frac{\alpha - 1}{\beta - 1} = k$, as expected for a system at criticality and experimentally verified in [40].

Therefore, the same critical value of the threshold, which gives the maximization of the fluctuation of the order parameter, also gives a critical avalanche distribution and universal scaling. This is in agreement with the picture discussed previously, showing that, at the critical threshold, there are intermittent reactivations of different stored patterns, which last for different durations, and the reactivation may be as large as the full network or involve only a short number of units. This suggests that the critical avalanches observed experimentally may be the manifestation of a system at the dynamic critical point of a phase transition, between a regime with replay of spatio-temporal dynamics patterns and a regime of non-replay.

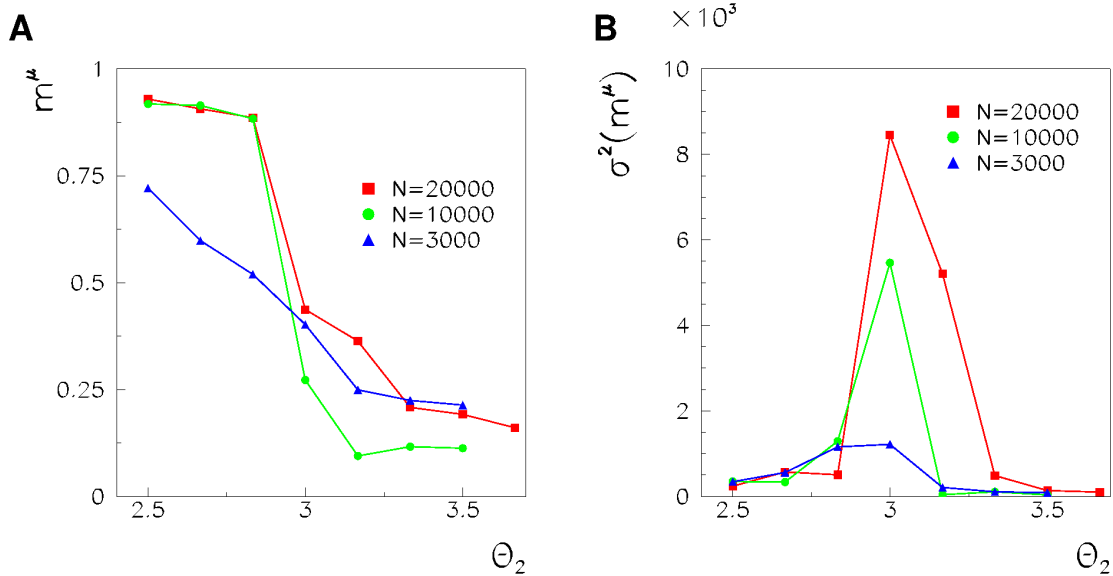

**Figure 4. Order parameter (A) and its fluctuations (B) as a function of $\Theta_2$, for $N=3000$ (triangles), 10000 (circles) and 20000 (squares).** The lower threshold is $\Theta_1=0.8$. Both the order parameter and its fluctuations show the signature of a phase transition at $\Theta_2=3.0$.
doi:10.1371/journal.pone.0064162.g004

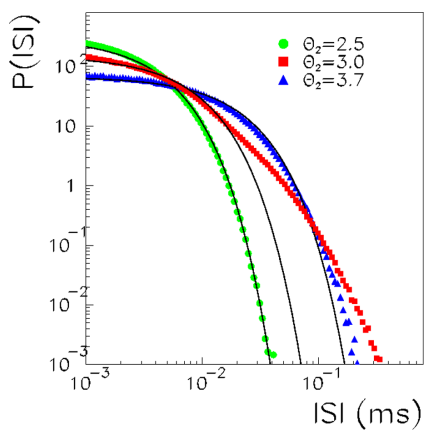

**Figure 5. The distribution of interspike intervals, between network spikes, is shown for $N=20000$, $\Theta_1=0.8$, and $\Theta_2=2.5$ (circles), 3.0 (squares) and 3.7 (triangles).** While the network ISI distributions at low and high values of $\Theta_2$ are quite well fitted by an exponential (shown as a solid black line), the network ISI at the critical threshold cannot be described by a single exponential (it strongly deviates from the exponential at intervals larger than 0.03 ms).
doi:10.1371/journal.pone.0064162.g005

## Discussion

We studied the spontaneous temporal dynamics in a noisy coupled network of spiking integrate-and-fire neurons, whose connectivity is designed in such a manner as to favor the spontaneous emergence of collective oscillatory spatio-temporal patterns of spikes. We introduce an order parameter to measure the overlap between the spintaneous collective dynamics and the stored phase-coded patterns, and we find a critical transition from a region of non-replay to a region of replay of the stored patterns.

At a critical value of the excitability, that is, of the spiking threshold $\Theta_2$, the system has a transition from a regime of Poissonian noise activity to a regime of spontaneous persistent replay of one of the stored spatio-temporal patterns. Exactly at the transition, the network spontaneous dynamics shows an intermittent reactivation of the stored patterns, with durations and sizes distributed over many scales. This suggests a relationship with the well-known phenomena of neural avalanches [5,6,11] observed in spontaneous cortical activity. Indeed, at the critical point, we observe avalanches whose size and duration distributions are power laws.

A model for neural avalanches related to the directed percolation model has been proposed recently [5]. Our model is different in that it makes use of spiking integrate and firing units and, more importantly, because synaptic connectivity is not random but has a structure derived from the learning rule defined in Eq. 9. The structure of the connectivity is responsible for the spatio-temporal correlations of the collective reactivations, which appear intermittently during the spontaneous dynamics. Notably, experimentally, the neural avalanches are related to the existence of repeated precise spatio-temporal activity patterns [4,11,13], and, to our knowledge, our model is the first one able to account for the recurrence of precise spatio-temporal patterns and give insight into the conditions under which these patterns can be retrieved.

To characterize the collective dynamics in the diverse regimes, we introduce in Eqs. (1) and (3) an order parameter. We see that, at the transition from low- to high-order parameter regime, the fluctuations of the order parameter are maximized. As we increase the size of the system, the order parameter transition becomes steeper and fluctuations more peaked, as expected in a continuous phase transition.

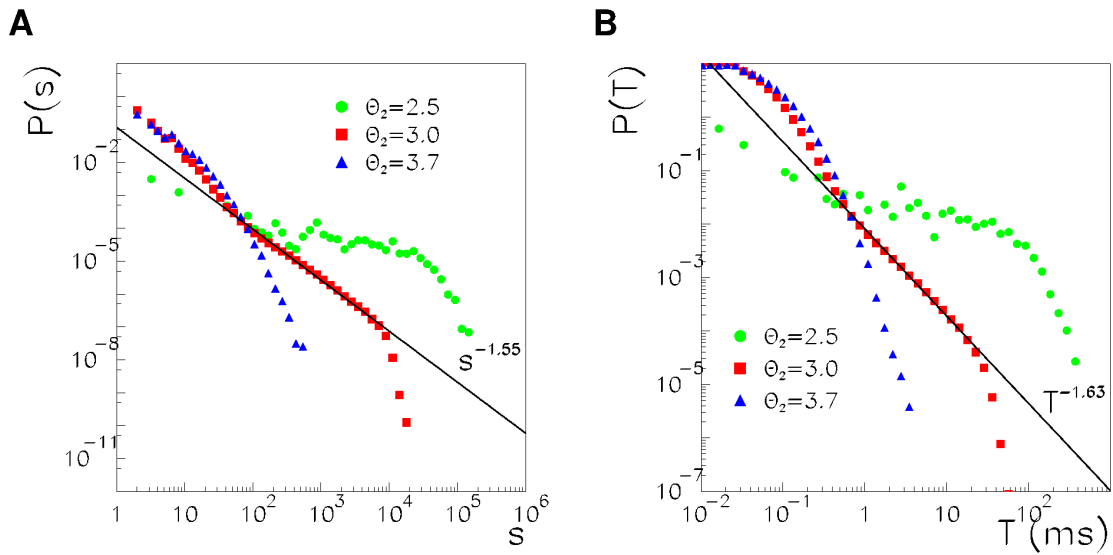

**Figure 6. Avalanche size (A) and duration (B) distributions are shown for the network with $N = 20000$ $\Theta_1 = 0.8$, and $\Theta_2 = 2.5$ (circles), 3.0 (squares) and 3.7 (triangles).** Solid lines are power law best fits of the $\Theta_2 = 3.0$ data in the intervals $10^2 < s < 10^4$ for the sizes and 1 ms $< T <$ 23 ms for the durations and give exponents $\beta = 1.55$ for the sizes and $\alpha = 1.63$ for the durations.
doi:10.1371/journal.pone.0064162.g006

The order parameter we introduce is a sort of extension of the Kuramoto order parameter, $r(t) \propto |\sum_j e^{i\phi_j(t)}|$, used in [5], which measures the synchrony of activity. However, the Kuramoto order parameter $r(t)$ measures only the overlap with a pattern having all phases $\phi^\mu$ equal, and therefore, the antiphase locking or, in general, the phase locking with different phases will reduce the order parameter. On the contrary, our order parameter measures

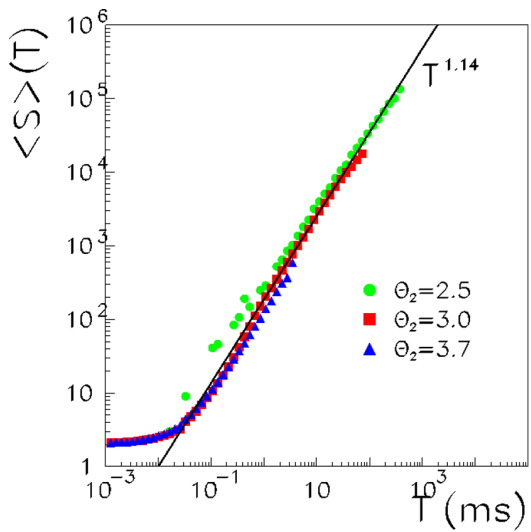

**Figure 7. Mean avalanche size as a function of the duration for the network with $N = 20000$ $\Theta_1 = 0.8$, and $\Theta_2 = 2.5$ (circles), 3.0 (squares) and 3.7 (triangles).** The solid line is a power law best fit of the $\Theta_2 = 3.0$ data in the interval 1 ms $< T <$ 23 ms and gives exponent $k = 1.14$.
doi:10.1371/journal.pone.0064162.g007

the similarity with a pattern having arbitrary phases $\phi_j^\mu$. In our model, we know *a-priori* the phases $\phi_j^\mu$ of the patterns, while experimentally the phases should be extracted from the data looking at the more repetitive patterns of phases. Interestingly, the onset of synchrony, together with the peak of synchrony fluctuations, observed in cortical cultures by [5] strongly suggests that the system undergoes a critical phase transition. This is in line with our model that shows the onset of order parameter $m$ together with the maximization of order parameter fluctuations and a power law in avalanche size and duration distributions.

The effects of noise in a cortical model was already addressed in an analogous Cowan-Wilson network model [41], with a connectivity structure similar to the one used here, and the existence of a stochastic resonance phenomenon was pointed out. It was shown that, in a particular regime of parameters, noise induces aperiodic collective synchronous oscillations, which mimics experimental in vitro cortical observations [42,43]. Under some conditions [41,44], the energy distribution of activity over low frequencies has a broadband with a power law decay, which indicates the existence of positive long-range time correlations in the sequences of bursts, as observed experimentally by Segev et al. [42,43]. The analog cortical model [41], equipped with a proper STDP learning rule, has been suggested to account for the spontaneous collective theta rhythms and the theta phase precession in hippocampus [34], while the coexistence of multiple patterns [45] and multiple frequency rhythms has then been addressed in [46]. The model introduced in [41] has then been extended to the case of binary units [47,48] and spiking IF units [28,49,50], but the study has been limited to the study of the replay dynamics induced by a short cue stimulation. In this paper, we study for the first time the dynamics, in the absence of cue stimulation, in a coupled network of IF units, and we show that intermittent collective emergence of multiple spatio-temporal patterns arises in presence of noise. For the first time, the existence of a nonequilibrium phase transition in the IF model has been pointed out, and the critical point has been characterized in terms of an order parameter and its fluctuations and power laws in avalanche size and duration distributions.

This model makes a strong connection between the evoked dynamics, induced by a cue sensory stimulation, and the spontaneous dynamics, in the absence of any sensory stimulus. Indeed, we found that the spatio-temporal patterns imprinted in the connectivity, which can be evoked by a cue stimulation [28], are the same spatio-temporal patterns that are intermittently reactivated by noise in the spontaneous dynamics at the critical point, as shown here. Recently, there was renewed interest in reverberatory activity [15] and in cortical spontaneous activity [51], whose spatio-temporal structure seems to reflect the underlying connectivity. Interestingly, the similarity between spontaneous and evoked cortical activities has been experimentally shown to increase with age [52] and with repetitive presentation of the stimulus [53].

## Models

We consider a recurrent neural network with $N(N-1)$ directed connections $J_{ij}$, where $N$ is the number of neural units. The single neuron model is a LIF model [30]. We use the spike response model formulation [30,31] of the LIF model. In this formulation, the postsynaptic membrane potential of neuron $i$ is given by

$$h_i(t) = \eta_i(t) + \sum_j J_{ij} \sum_{\hat{t}_j > \hat{t}_i} \epsilon(t - \hat{t}_j), \qquad (4)$$

where $\eta_i(t)$ is a Poissonian noise, $J_{ij}$ are the synaptic connections, $\epsilon(t)$ describes the response kernel to incoming spikes, and the sum over $\hat{t}_j$ runs over all presynaptic firing times following the last spike of neuron $i$. Namely, each presynaptic spike $j$, with arrival time $\hat{t}_j$, is supposed to add to the membrane potential a postsynaptic potential of the form $J_{ij}\epsilon(t-\hat{t}_j)$, where

$$\epsilon(t - \hat{t}_j) = K\left[\exp\left(-\frac{t-\hat{t}_j}{\tau_m}\right) - \exp\left(-\frac{t-\hat{t}_j}{\tau_s}\right)\right] \Theta(t - \hat{t}_j) \qquad (5)$$

where $\tau_m$ is the membrane time constant (here 10 ms), $\tau_s$ is the synapse time constant (here 5 ms), $\Theta(t)$ is the Heaviside step function, and K is a multiplicative constant chosen so that the maximum value of the kernel is one. The sign of the synaptic connection $J_{ij}$ sets the sign of the postsynaptic potential change.

A Poissonian noise $\eta_i(t)$, related to the spontaneous neurotransmitter release at individual synapses, as well as other sources of inhomogeneity and randomness that determine an irregular background synaptic noise in vitro, is modeled as

$$\eta_i(t) = \sum_{\hat{t}_{\text{noise}} > \hat{t}_i} J_{\text{noise}} \epsilon(t - \hat{t}_{\text{noise}}). \qquad (6)$$

The times $\hat{t}_{\text{noise}}$ and the strengths $J_{\text{noise}}$ are extracted randomly and independently for each neuron $i$. The intervals between times $\hat{t}_{\text{noise}}$ on the single neuron are extracted from a Poissonian distribution $P(\delta t) \propto e^{-\delta t/\tau_{\text{noise}}}$, while the strength $J_{\text{noise}}$ is extracted for each time $\hat{t}_{\text{noise}}$ from a Gaussian distribution with mean $\bar{J}_{\text{noise}}$ and standard deviation $\sigma(J_{\text{noise}})$. In all simulations noise is given by Eq. (6) with $\tau_{noise}=1$ ms, $\bar{J}_{noise}=0$, and $\sigma(J_{noise})=0.2$.

When the membrane potential $h_i(t)$ exceeds the spiking threshold $\Theta^i$, a spike is scheduled, and the membrane potential is reset to the resting value of 0. No refractory period is taken into account. While in previous work [28] we used a unique value of the spiking threshold $\Theta^i$ for all units, here we use two values of $\Theta^i$, a low threshold $\Theta_1$ for $N_1 < N$ units more sensible to noise and a higher threshold $\Theta_2$ for the other $N_2 = N - N_1$ units. We form the hypothesis that, due to the many sources of inhomogeneity and randomness, for each stored pattern there is a subset of units, with consecutive phases in the pattern, that have a low threshold $\Theta_1$. These low-threshold units will be more sensitive to noise, while the units with a higher threshold $\Theta_2$ will be activated mainly only when the collective replay of the pattern has emerged. In all simulations a fraction equal to 3.3% of the neurons has a threshold $\Theta_1 = 0.8$, while the other units have a threshold $\Theta_2$, which has a different value ranging from 2.5 to 3.7.

Numerical simulations of the dynamics are performed for a network with $P$ stored patterns, where connections $J_{ij}$ are determined via the learning rule, inspired by the STDP, previously introduced in [28,32–34,54]. The connections $J_{ij}$ are designed during the learning mode. After the learning stage, the connection values are frozen, and the spontaneous collective dynamics are studied. During the learning stage, the average change in the connection $J_{ij}$, occurring in the time interval $[-T,0]$, due to the presentation of a periodic spike trains of period $T^\mu$ is formulated as follows:

$$\delta J_{ij} = \frac{1}{N}\frac{T^\mu}{T} \int_{-T}^{0} dt \int_{-T}^{0} dt' \, x_i(t) A(t-t') x_j(t') \qquad (7)$$

where $1/N$ and $T^\mu/T$ are normalization factors, $x_j(t)$ is the activity of the presynaptic neuron at time t, and $x_i(t)$ the activity of the postsynaptic one. In the STDP, the learning window $A(\tau)$ is the measure of the strength of the synaptic change when a time delay $\tau$ occurs between pre- and postsynaptic spikes. Here, the patterns to be stored are defined as precise periodic sequence of spikes, i.e., phase-coded patterns. When pattern $\mu$ is replayed, the activity of neuron $j$ is periodic, with spikes at times $t_j^\mu$,

$$x_j^\mu(t) = \sum_{n=-\infty}^{\infty} \delta\left[t - (t_j^\mu + nT^\mu)\right], \qquad (8)$$

where $t_j^\mu + nT^\mu$ is the set of spikes times of unit j in the pattern $\mu$ with period $T^\mu$.

From Eqs. (7) and (8), when the learning time is longer than the period $T^\mu$ of the learned pattern, we have

$$\delta J_{ij}^\mu = \frac{1}{N} \sum_{n=-\infty}^{\infty} A(t_j^\mu - t_i^\mu + nT^\mu). \qquad (9)$$

The window $A(\tau)$ is the one introduced and motivated by [55], $A(\tau) = a_p e^{-\tau/T_p} - a_D e^{-\eta\tau/T_p}$ if $\tau > 0$ and $A(\tau) = a_p e^{\eta\tau/T_D} - a_D e^{\tau/T_D}$ if $\tau < 0$, with the same parameters used in [55] to fit the experimental data of [36], $a_p = \gamma[1/T_p + \eta/T_D]^{-1}$ and $a_D = \gamma[\eta/T_p + 1/T_D]^{-1}$ with $T_p = 10.2$ ms and $T_D = 28.6$ ms, $\eta = 4$, $\gamma = 42$.

This function satisfies the balance condition $\int_{-\infty}^{\infty} A(\tau)d\tau = 0$. Notably, when $A(\tau)$ is used in Eq. (9) to learn phase-coded patterns with uniformly distributed phases, then the balance condition assures that the sum of the connections on the single neuron $\sum_j J_{ij}$ is of order $1/\sqrt{N}$, and therefore, it assures a balance between excitation and inhibition. Note that, as we are studying a network of excitatory neurons, the negative connections have to be thought as connections mediated by fast inhibitory interneurons.

The spike patterns used in this work are periodic spatio-temporal sequences made up of one spike per cycle, each of which has a phase $\phi_j^\mu$ randomly chosen from a uniform distribution in $[0,2\pi)$. In each pattern, information is coded in the precise time delay between spikes of unit $i$ and unit $j$, which corresponds to a precise phase relationship among units $i$ and $j$. A spatio-temporal pattern represented in this way is often called a phase-coded pattern. A pattern's information is coded in the spiking phases, which, in turn, shape the synaptic connectivity responsible of the emerging dynamics and the memory formation. The set of timing of spikes of unit $j$ is defined as

$$t_j^\mu + nT^\mu = \frac{1}{2\pi\nu^\mu}(\phi_j^\mu + 2\pi n).$$

Thus, each pattern $\mu$ is characterized by the frequency $\nu^\mu = 1/T^\mu$ and the specific phases of spike $\phi_j^\mu$ of the neurons $j=1,\ldots,N$. In all simulations we use $\nu^\mu = 3$ Hz, and randomly extracted phases $\phi_j^\mu$. When multiple phase-coded patterns are stored, the learned connections are simply the sum of the contributions from individual patterns, namely,

$$J_{ij} = \sum_{\mu=1}^{P} \delta J_{ij}^\mu. \qquad (10)$$

## Author Contributions

Analyzed the data: SS ADC. Wrote the paper: SS ADC.